\documentclass[12pt]{article}\usepackage[hyperfootnotes=false]{hyperref}
\usepackage{epsfig}
\usepackage{float}
\usepackage{empheq}
\usepackage{bbold}

\usepackage{xspace}

\makeatletter

\@addtoreset{equation}{section}
\makeatother

\usepackage[utf8]{inputenc}
\usepackage{amsmath}
\usepackage{caption}
\usepackage{amsmath}
\usepackage{amssymb}
\usepackage{graphicx}

\newlength{\abstractwidth}
\renewcommand{\thefootnote}{\fnsymbol{footnote}}
\renewcommand{\thanks}[1]{\footnote{#1}} 
\newcommand{\starttext}{
\setcounter{footnote}{0}
\renewcommand{\thefootnote}{\arabic{footnote}}}

\newcommand{\be}{\begin{equation}}
\newcommand{\bea}{\begin{eqnarray}}
\newcommand{\eea}{\end{eqnarray}}
\newcommand{\beq}{\begin{equation}}
\newcommand{\ee}{\end{equation}}

	\newcommand*\widefbox[1]{\fbox{\hspace{2em}#1\hspace{2em}}}

	\def\dss{de Sitter space}
	\def\dsp.{de Sitter space.}
	\def\eq{&=&}

	\def\la{\langle}
	\def\ra{\rangle}
	\def\simleq{\; \raise0.3ex\hbox{$<$\kern-0.75em
			\raise-1.1ex\hbox{$\sim$}}\; }
	\def\simgeq{\; \raise0.3ex\hbox{$>$\kern-0.75em
			\raise-1.1ex\hbox{$\sim$}}\; }
	
	\def\bi{\begin{itemize}}
		\def\ei{\end{itemize}}

	\def\dof{degrees of freedom }

	\def\CH{{\cal{H}}}

	\def\CT{{\cal{T}}}

	\def\CZ{{\cal{Z}}}

	\def\Tr{\rm Tr \it}
	\def\bsub{ \begin{subequations}
			\begin{empheq}[box=\widefbox]{align}  }
			\def\esub{ \end{empheq}
	\end{subequations}}
	
	\def\1{\(  \mathbb{1} \)}

	\def\lf{\left(}
	\def\rg{\right)}

	\def\bn{\bigskip \noindent}

	\def\dk{${\rm DSSYK_{\infty}}$}
	
	\makeatletter
	\g@addto@macro\normalsize{%
		\setlength\abovedisplayskip{10pt}
		\setlength\belowdisplayskip{20pt}
		\setlength\abovedisplayshortskip{10pt}
		\setlength\belowdisplayshortskip{20pt}
	}
	\makeatother
	
	\usepackage{color}

	\usepackage{authblk}
	\title{\Large \bf More About the Spontaneous Breaking of  Time Reversal in de Sitter Space }

	\author[1,2]{\Large Leonard Susskind}
	\affil[1]{LITP and Department of Physics, Stanford University, Stanford, CA 94305-4060, USA \vspace{1em}}
	\affil[2]{Google, Mountain View, CA, USA}
	\date{}
	\usepackage{upgreek}

\begin{document}
		
		\begin{titlepage}
			\maketitle
			
			\begin{abstract}
			\Large
			
			It is widely thought that the quantum theory of de Sitter space requires the existence of a physical observer in the static patch. What exactly is meant by an observer is unclear; it could be anything from a few photons with energy just above the Gibbons-Hawking temperature to a gravitationally bound cluster of galaxies.
			In a recent note  I explained that one way the need for observers  can  arise is from the spontaneous breaking of time-reversal symmetry. This longer paper expands on the subject,  filling in conceptual arguments  that were implicit but not explicitly stated in the earlier paper.

 \end{abstract}

		\end{titlepage}
		
		\rightline{}
		\bigskip
		\bigskip\bigskip\bigskip\bigskip
		\bigskip
		
		\starttext \baselineskip=17.63pt \setcounter{footnote}{0}

	\LARGE

		\tableofcontents
\section{Introduction}\label{intro}

There is a widespread belief that the quantum mechanics of de Sitter space only makes sense  if there is an observer present
 \cite{Chandrasekaran:2022cip}.  Discrepancies between semiclassical
 gravity and static patch holography have been resolved by introducing a ``feature"   localized at the pode of the static patch. According to the terminology of   
 \cite{Chandrasekaran:2022cip}\cite{Abdalla:2025gzn}\cite{Ivo:2025yek}\cite{Chen:2025jqm} this feature is called an observer, although  it may have little to do with conventional observation by almost classical apparatuses. 
 
 In this paper I will be concerned with one particular discrepancy. According to static patch holography  two-point correlation functions in \dss \ are real; the imaginary parts vanish   \cite{Susskind:2023rxm}. This is in contrast to  bulk field theory in a de Sitter background where the two-point functions are complex. 
  The ability of a  observer, localized at the pode,  to resolve this kind of  discrepancy  is surprising because the semiclassical calculation may involve bulk \dof \ localized extremely far from the pode---for example near the horizon---and yet no matter how far away the pode is from  the region of the semiclassical calculation, the presence of an observer at the pode changes things enough to resolve the discrepancy. On the face of it this is a form of non-locality forbidden in bulk quantum field theory.

Generally the literature on this subject does not spell out exactly  what is meant by an observer. In fact it seems that it could be anything from a galaxy to a tiny out-of-equilibrium feature in the Hawking radiation---no more than a handful of photons with energy of order  the Gibbons-Hawking temperature. Calling such a feeble phenomenon  an observer seems misleading to me; I prefer  feature  to the overly suggestive observer. It is important to keep in mind that feature is just a name for an out-of-equilibrium fluctuation that could be almost anything.

 Features comprise a more general class than observers. Observers, among other requirements, are heavy enough to be localized near the pode; Features can be heavy, i.e., they can be observers, or they can  much lighter and can have wavelengths of cosmic scale. Localized observers will be useful for arguing that there is a breakdown of cluster-decomposition in \dss, but almost any kind of feature is sufficient to resolve the mismatch.
 
  One thing about observers or features  is that they should have some form of time-reversal asymmetry to distinguish future and past. I'll call such an asymmetric feature a ``clock."

Now  let's turn to the region $h$ in figure \ref{lattice}. I'll refer to it as the ``laboratory." It is the region of interest where observable phenomena  take place, the phenomena being described by a set of correlation functions localized near $h.$

In a recent  note  \cite{Susskind:2025wyn} I argued that the need for an observer (or feature)  has nothing to do with the consistency of quantum mechanics; it is due to a spontaneous  breaking (SSB) of time-reversal symmetry. The  feature  kicks off an instability of the symmetric vacuum and restores cluster decomposition\footnote{Violations of cluster decomposition, aka infinite range correlations, are generally associated with spontaneous symmetry breaking.}. 

In this followup the arguments of \cite{Susskind:2025wyn} will be clarified and made more complete.

I will use the notation $\CT$ for time-reversal, both the symmetry and the antiunitary  time-reversal operator. It has the following properties:

$\CT$ is antiunitary. It is the product of a unitary operator and complex conjugation.

$\CT$ satisfies $$\CT^{-1} = \CT.$$

I'll begin by  reviewing  the conclusions of \cite{Susskind:2025wyn}:
\begin{enumerate}
\item There is a breakdown of bulk cluster decomposition in the usual de Sitter vacuum. In the bulk theory the de Sitter vacuum means  the Hartle-Hawking state;   
in the holographic boundary theory it is  the maximally  entangled thermofield-double state.

\item A failure to cluster is symptomatic of spontaneous symmetry breaking. For the case at hand the spontaneously broken symmetry is $\CT$ (time-reversal).

\item
 When SSB takes place the symmetric vacuum is unstable to very slight perturbations which can kick the vacuum into one of the stable asymmetric states. 

\item In the unstable $\CT$-symmetric vacuum correlation functions do not match semiclassical expectations; for example two-point functions are necessarily real in the symmetric vacuum but are complex in the semiclassical theory \cite{Susskind:2023rxm}.
In the stable asymmetric vacuum correlation functions are consistent with semiclassical expectations.

\item

I will add one more thing to the list which may or may not have been clear from \cite{Susskind:2025wyn}. There is a subtlety about the meaning of time-reversal in de Sitter space; in ordinary quantum mechanics clocks may be thought of as registering  absolute Newtonian time; for that reason we never need to introduce them explicitly. They live \it outside the system of interest \rm and do not interact with it\footnote{If the clock interacted with the system the system would back-react on the clock and the clock could not register absolute time.}.

In a closed universe without boundaries, e.g., de Sitter space,
the only physically meaningful time is that  registered by a real material clock. This raises a question about the meaning of $\CT$: 

\bn
\it When we reverse time do we only reverse the motion of the system, or do we also reverse the clock? 
\bn
\rm

In ordinary QM we reverse only the system but not the  clock. In de Sitter space we must also reverse the clock. I will  explain the subtlety by example in what follows.

Let me depart from the technical argument here and give an analogy. Imagine a quantum movie of a quantum oscillator and a real clock in the movie theater keeping time. Now run the movie backward. Is there really a symmetry or do we also have to run the clock backward? In principle these are two different operations. If the clock and the system are totally non-interacting there is no important difference between the two but if there are any interactions then only the combined time reversal is a symmetry.

\end{enumerate}

 Harlow and collaborators have argued that in quantum gravity $\CT$ is a gauge symmetry \cite{Harlow:2023hjb}. I'll  discuss this issue, but in short I think it is important  to consider it so. 

In the next two sections I will illustrate some basic concepts about $\CT$ and SSB using elementary examples. After that I will put them together in the context of \dss.

\section{Time-Reversal With or Without Clocks} \label{withor}

Let's begin with a system and a clock. I'll choose a very simple system: the Harmonic oscillator and later a very simple clock.
\subsection{The System} \label{system}
The equations are too well known to require explanation, I'll just write them down for the case $\omega=1.$
\bea 
[a^-, a^+]\eq 1  \cr\cr
H_{\sigma}&=&a^+a^- \cr\cr
 x(t) \eq    \frac{a^+e^{it}  + +a^-e^{-it}}{2}  
\label{x(t)}
\eea
The subscript $\sigma$ means system.

By the usual definition of time-reversal the harmonic oscillator is  a $\CT$-invariant system. Let's test that out.
The operator $C$ 
\bea 
C \eq [x(-s),x(s)]  \cr \cr
x(s) \eq e^{iHs}xe^{-iHs} 
\label{C(t)}
\eea
(for any fixed\footnote{A more general version of \eqref{C(t)}
is $C(s,t) = [x(t-s), x(t+s)]$
} 
numerical value of $s$) changes sign under $\CT$; in other words it is $\CT$-odd.
To see that we use $$\CT x(t)\CT =x(-t)$$  and we find
$ \CT C \CT= -C.$
Now consider the expectation value of $C$ in any $\CT$-invariant state, for example the oscillator ground state. Since $C$ is odd and the state is even under $\CT$ we might expect $$\langle 0 | C|0\ra =0.$$ But explicit calculation gives,
\be  
\la0| C|0\ra = \frac{i}{2} \sin{s}
\label{sint}
\ee
Only the real part of $\la0| C |0\ra $ vanishes for any $s\neq 0.$  Evidently the logic which says that  $\la0| C |0\ra $ vanishes  is wrong.

\subsection{The Clock} \label{clock}

In ordinary quantum mechanics time is ``Newtonian," that is to say  time recorded by an ideal clock that lives outside the system without interacting with it. We will denote it by $t.$  In the real world, especially in cosmology, any clocks that exist are part of the universe, the universe being
 the system we are studying. This is not important in ordinary quantum mechanics but in a closed universe like \dss \ it is all important.

Let us introduce a simple model for a clock, namely a free particle with coordinate $\tau$ and conjugate momentum $\pi.$ 
\be 
[\tau, \pi] = i
\label{[tp]=i}
\ee

The clock Hamiltonian is,
\be  
H_c=|\pi|
\ee
The $\tau$ axis is the ``face of the  clock" and the $\tau$ operator represents  clock time. Ideal Newtonian  time is still denoted by  $t.$

Consider a state of the clock with momentum-space wave function supported entirely in  positive $\pi.$ In $\tau$-space the wave packet will move rigidly  toward larger $\tau$ without changing shape. For such wave packets we can identify the clock time with Newtonian time. 
Such a clock is a ``forward-going--clock" (FGC).
 \be  
 \tau = t  \ \ \ \ \ \text{FGC}
 \label{fgc}
 \ee

Now let us combine the clock with the system and replace Newtonian time $t$ with clock time $\tau$ in the various oscillator equations. Since $\tau =t$ for our FGC clock this is trivial (just replace $t$ by $\tau$). In particular   equation \eqref{sint} becomes
\be  
\la0| C |0\ra = \frac{i}{2} \sin{\tau}
\label{sintauf}
\ee

Now let's consider a time-reversed ``backward-going-clock" with momentum space wave function supported in the negative $\pi$ region. For this BGC,
\be  
\tau = -t  \ \ \ \ \ \text{BGC}
\ee

Equation \eqref{sintauf} is replaced by,
\be  
\la0| C(\tau) |0\ra = -\frac{i}{2} \sin{\tau}
\label{sintaub}
\ee

Neither of these two states, $$\text{system} \otimes \text{FGC}$$ and 
$$\text{system} \otimes \text{BGC}$$ are $\CT$ invariant. Only the superposition of them is $\CT$-invariant. Calling such a $\CT$invariant state 
$|\psi \ra$ we may write it in the form,

\be 
|\psi\ra = |0\rangle \otimes \big(  \ |FGC\ra + |BGC\ra \ \big).
\label{psi}
\ee
The state $|\psi\ra$ is invariant  under the  full time-reversal in which both the system and the clock are reversed. 

It is easily seen that
\be 
\la \psi|C |\psi \ra =0.
\label{expc=0}
\ee
In the fully time-reversed state in which both the system and the clock are reversed the expectation value of the $\CT$-odd operator in the $\CT$-even state is indeed zero.

Let me emphasize that it is not enough to time reverse the clock together with the system but we also need to replace ideal time $t$ by the clock time 
$\tau$  (see  \cite{Chandrasekaran:2022cip}).

 \subsection{Time Reversal of the Clock}

One might think that under $\CT$ the clock time $\tau$ should be odd but that's incorrect. $\tau$ is the position of a particle or the hand of a clock. As such it is $\CT$ invariant
\footnote{In the quantum mechanics of a particle  the position coordinates are invariant under time-reversal and the momentum changes sign.}.

\be 
\CT   \tau    \CT  = \tau.
\ee

The momentum $\pi$ is odd,
\be 
\CT \pi \CT = - \pi
\ee
and so is 
the velocity of the clock,
\bea 
\dot{\tau} \eq  -i \big[ \tau, H \big] \cr \cr
\eq  -i\big[\tau, |\pi| \big] \cr \cr
\eq \frac{\pi}{|\pi|}.
\label{vel}
\eea

\subsection{Projection Operators}

Before going on I want to define some notation which will come in handy when we apply these lessons to de Sitter space. Let $\Pi_f$ be a projection operator onto states with a FGC. In other words it projects onto states with positive values of $\pi$ while doing nothing to the system. Similarly $\Pi_b$ projects onto states with a BGC---negative values of $\pi$. 
The sum  $\Pi_f+\Pi_b$ is just the unit operator. It is obviously $\CT$ symmetric.

We also introduce the difference operator $\Pi_-,$
\be
\Pi_f-\Pi_b = \Pi_-
\label{Pipm}
\ee
 $\Pi_-$  is a difference of two non-overlapping projection operators.
It  is odd   under $\CT$.

  Using projection operators we can write equations \eqref{sintauf}  and  \eqref{sintaub} in the compact form,
  \be  
\la C\ra = \frac{i}{2}  \Pi_-  \sin{\tau}
\label{sintauf}
              \ee
Since $\Pi_-$ is odd under $\CT$ its expectation, as well as the expectation value of $C$ vanishes  in any $\CT$-invariant state.

To summarize: There are two versions of time-reversal; one in which we reverse the system but not the clock, the second in which we reverse everything including the clock. In the first version $\la C \ra \neq 0,$ in the second $\la C \ra = 0$ for any $\CT$-invariant state.

\section {Spontaneous Symmetry Breaking}

\begin{figure}[H]
		\begin{center}
			\includegraphics[scale=.2]{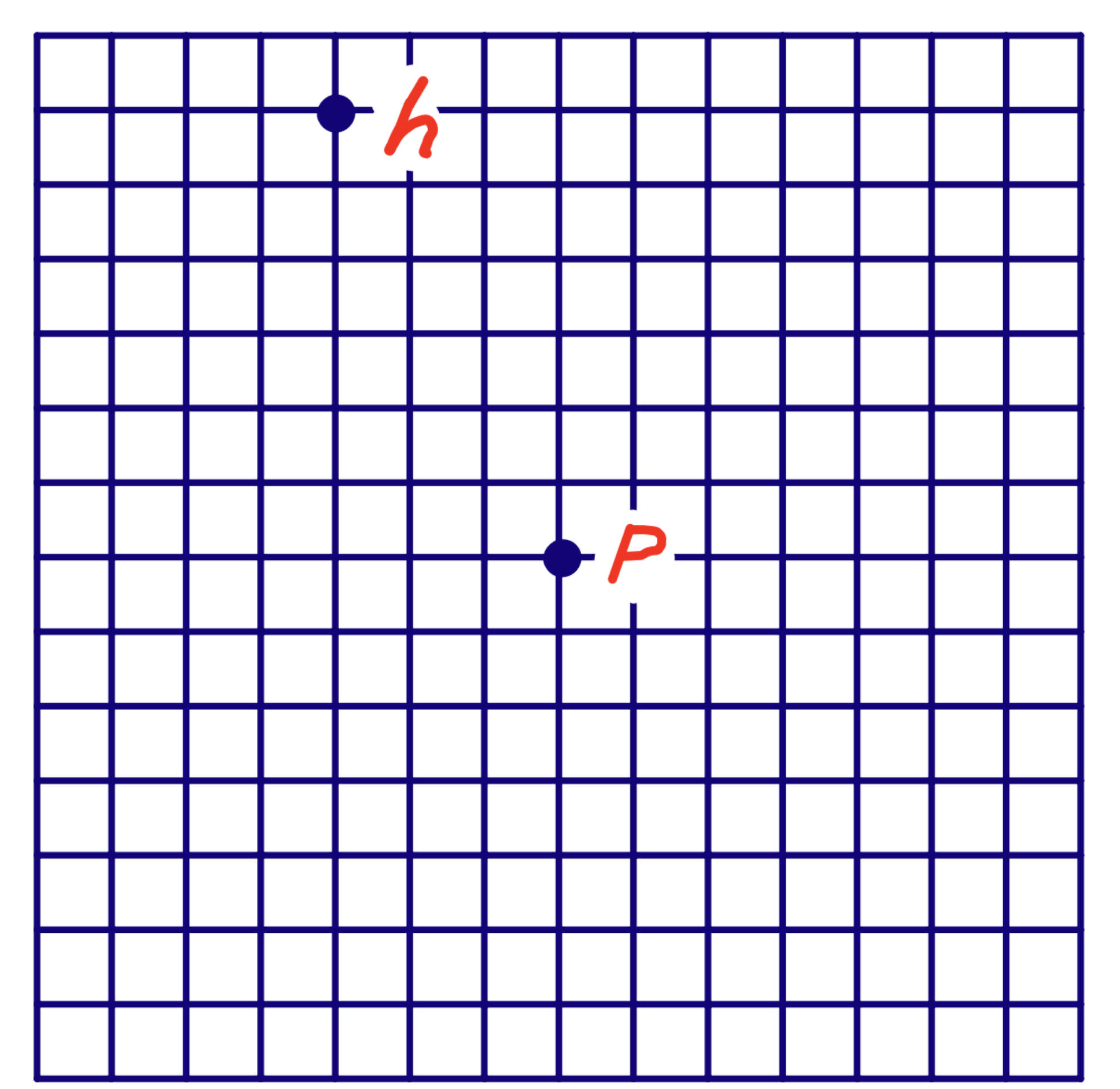}
			\caption{}
			\label{lattice}
		\end{center}
	\end{figure}

I will illustrate spontaneous symmetry breaking   by the simple example of the 2D Ising model.  Take a very large $N\times N$ lattice ($N>>1$) populated by spins shown in figure \ref{lattice} for $N=15.$  
The lattice is the analog of the static patch of \dss; the center of the lattice playing the role of the pode and the boundary  playing the role of the horizon.

The Hamiltonian is 
\be  
H = \sum_{\text{nn}} Z_iZ_j +\epsilon \sum_i X_i  \ \ \ \ \ (\epsilon<<1)
\label{H}
\ee
where $\sum_{nn}$ means sum over nearest neighbors. The Hamiltonian is invariant  under the discrete $\CZ_2$ symmetry $K,$
\bea 
K Z_i K \eq -Z_i  \cr  \cr
KX_iK &=& X_i   \cr  \cr
KY_i K \eq -Y_i.
\label{K}
\eea
The eigenstates of of $Z$ will be labeled by $|u\ra$ and $|d\ra$ (up and down). By definition,
\bea
Z |u\ra &=& |u \ra  \cr \cr
 Z |d \ra &=& - |d \ra 
 \label{eigen}
\eea

The ground state for $\epsilon <<1$ has the $\CZ_2$ invariant form of a ``cat state."
\be 
|gs\ra = |uuu\cdots u\ra  +  |ddd\cdots d\ra =|U\ra +|D\ra
\label{gs}
\ee

\subsection{Breakdown of Clustering}\label{brakclus}

Consider two very distant points 
$p \  \text{and} \ h$                                                                                                                                                                                                                                                                                                                                                                                                                                                                                                                                                                                                                 separated by a lattice distance $\sim N.$ To be definite let $p$ be at the center  and $h$ be out near the boundary  as in figure \ref{lattice}.

Call the spin at $p$  the ``observer" and the region near $h$ the ``laboratory," i.e., the place where experiments take place. The laboratory may well contain a genuine observer but that is incidental.

One  finds
\bea 
\langle gs|Z_p |  gs \ra \eq 0 \cr  \cr
\langle gs|Z_h |  gs \ra \eq 0 \cr  \cr                                                                                                                                                                                                                                                                                                                                                                                                                                                       
\langle gs|Z_p Z_h|  gs \ra \eq 1. 
\label{gszgs}
\eea
Given that $p  \  \text{and} \ h$ are arbitrarily far apart, equations \eqref{gszgs} imply a violation of cluster decomposition.
States that violate cluster decomposition such as the cat state \eqref{gs} are not normally seen in nature; they are highly unstable and tend to collapse, either to $|uuu\cdots u \ra$ or $|ddd\cdots d\ra.$ 

Restoring cluster decomposition can be done in a number of equivalent ways, all of which pick out the observer as a  reference point to which other spins can be compared (dressed). The first way is to introduce an addition to the Hamiltonian which only involves $p.$
\be  
H = \sum_{\text{nn}} Z_iZ_j +\epsilon \sum_i X_i  + bZ_p
\label{totalH}
\ee
 In the $N\to \infty$ limit   if $b$ remains finite, no matter how small,   it will have 
 a dramatic global effect on the ground state even though the perturbation only involves one spin. Depending on the sign of $b$ the new ground state will either be $$|U\ra=|uuu\cdots u\rangle$$ or $$|D\ra =|ddd\cdots d\rangle.$$ Both of these states individually satisfy cluster decomposition.
 
 The symmetric ground state \eqref{gs} is unstable to a tiny  perturbation which not only polarizes the observer  spin but the entire lattice.  For $N\to \infty$ the magnitude of the effect becomes independent of the magnitude of $b.$
 
 \subsection{Gauging $\CZ_2$} \label{gauge}
 
 Adding an extra symmetry breaking as in \eqref{totalH} is artificial device which has no analog in \dss. Another way to accomplish the same result---cluster decomposition---is to gauge 
 gauge $\CZ_2 $ by  restricting the physical space of states and the observables to be  $\CZ_2$ invariant.  As before we single out $p$ as the observer and ``dress" all observables to $p.$ This simply means replace all other
\footnote{In this context the index $i$ runs over all spins other than the observer $Z_p.$}
    $Z_i$ by the dressed operators $\bar{Z}_i,$
 \be  
 \bar{Z}_i = Z_p Z_i
 \label{dress}
 \ee 
 Correlations of the gauge invariant dressed operators satisfy cluster decomposition and are identical to the correlations in the state
 $|U\ra. $

 \subsection{Projection Operators}
 Let us introduce projection operators $\Pi_u$ and $\Pi_d$ which project onto up and down states of the ``observer."  They satisfy the following identities,
 \bea 
 \Pi_u^2 \eq \Pi_d^2 =1 \cr  \cr
 \Pi_u \Pi_d \eq 0 \cr  \cr
 \Pi_u + \Pi_d \eq 1
 \label{Ps}
 \eea
 They also act as the identity on all spins other than $p.$
 
 We also introduce the $\CZ_2$-odd operator,
 \be 
 \Pi_- =  \Pi_u - \Pi_d 
 \label{P-}
 \ee

 \bn
 
Note that $\langle gs | Z_i |gs \ra =0$ but   $\langle gs | Z_i \Pi_-|gs \ra \neq 0.$

\section{Time-reversal in de Sitter Space}
We will now apply the lessons of the previous sections to de Sitter Space, in particular to the paradox  described in  \cite{Susskind:2023rxm}. 

\subsection{Static Patch Holography}

We assume that a de Sitter static patch is described by a holographic system localized on a stretched horizon about  a Planck distance from the mathematical horizon. The basic setup is a set of discrete  holographic degrees of freedom, a finite dimensional Hilbert Space $\CH$, a Hamiltonian $H$, and a maximally mixed (infinite Boltzmann temperature) density matrix. The entropy is the logarithm of the dimension of the Hilbert space. Expectation values of operators are normalized traces,
\be 
\la A \ra = \Tr A
\label{expA}
\ee 
The \dof \ are in static thermal equilibrium and the only things that happen are thermal fluctuations.

The Hamiltonian is assumed to be time-reversal invariant.

Also assume an approximate dictionary relating the holographic \dof \ to bulk field \dof \  $A(r,  \Omega, t).$  (Here  $r,   \Omega$ and $t$ are the usual static patch radial, angular and time coordinates.)

\subsection{Clocks in the  Static Patch}

In the simple model of section \ref{withor} we had the option of defining time reversal with or without reversing the clock. In \dss \ we don't have that option; clocks are material systems composed of the same \dof \ that comprise everything else. Moreover they interact with everything else, if no other way, gravitationally. The concept of a time-reversal operation that only acts on the system, and not on the clock is at best an imprecise one. From now on $\CT$ operates on all 
\dof \ including any clocks that happen to be present.

Clocks are ``lumps" of matter which we  assume are localized near the pode of a static patch (see however section \ref{vpoint}). Like any other features they are rare thermal fluctuations which are emitted by the  stretched horizon and are eventually reabsorbed. The probability to find a lump of mass $M$ is,
\bea 
\text{Prob}  \eq \exp{\lf -\frac{M}{T_{gh}}  \rg} \cr \cr
\eq            \exp{\lf  -2\pi \ell M  \rg}
\eea
where $\ell$ is the de Sitter radius and the subscript $gh$ refers to Gibbons-Hawking. 

It is evident that  lumps which can reasonably called clocks are extremely (exponentially) rare,  but one can condition of the existence of a clock. A formal way to do that is to introduce projection operators in the static patch Hilbert space. The projection operators are bulk operators that in principle can be expressed in terms of the holographic degrees of freedom. The details are not important other than from the bulk viewpoint they are localized near the pode, far from the horizon.

Let $\Pi$ be a $\CT$-symmetric projection operator that projects onto a subspace in which a clock of some particular type is present at the pode. I will assume that the clock can be either forward or backward going,  and that $\Pi$ is symmetrically composed of two non-overlapping projection operators: a forward-going and a backward going projection. We also define $\Pi_-,$
\bea 
\Pi \eq \Pi_+ = \Pi_f + \Pi_b \cr \cr
\Pi_f \eq \CT \Pi_b \CT   \cr \cr
\Pi_- \eq  \Pi_f - \Pi_b \cr \cr
\label{P=Pf+Pb}
\eea
Note that $\Pi_-$ is not a projection operator;  its square satisfies,
\be  
\Pi_-^2= \Pi \neq 1.
\label{p-2neq1}
\ee

As in section \ref{withor} we will use the notation $\tau$ for clock time. For a FGC $$\tau =t$$ and for a BGC $$\tau = -t.$$

Conditioning on the existence of a clock without regard to whether it is a FGC or BGC is accomplished by replacing the maximally mixed density matrix by 
$\Pi$. Expectation values are given by,
\be  
\langle A \ra = \Tr \Pi A = \Tr \big( \Pi_f + \Pi_b    \big) A
\ee

 We may also  condition on a FGC (BGC) by applying $\Pi_f$ ($\Pi_b$).
 
 \bea  
\langle A \ra \eq \Tr \Pi_f A  \cr \cr
\eq \frac{1}{2}  \Tr  \  \big( \Pi_-   A \big)
\eea

So far this parallels section \ref{clock}. But as a bonus it   also parallels section
 \ref{brakclus}; 
 the location of the operators $\Pi, \Pi_-,  \Pi_f,  \Pi_b$ 
  is at the pode---the center of
  the static patch analogous to the center of the square lattice in figure
 \ref{lattice}. The operator $A$ is out near the horizon analogous to the point $h.$ 
 
In parallel with \eqref{C(t)} we also consider the operator $$C = [A(-s), A(s)]$$ for a fixed numerical value of a time $s.$
$C$ is odd under $\CT.$ 

The expectation value of $C$ is an order parameter for SSB of time-reversal analogous to $Z_i$ being an order parameter for the $\CZ_2$ symmetry $K$ in the Ising model.  

Since $\Pi_- $ and $C$ are $\CT$-odd their expectation values vanish in the maximally mixed state,
\bea 
\Tr \Pi_- \eq 0 \cr \cr
\Tr C \eq 0
\eea

 The distance between the pode and the horizon becomes infinite in the limit of infinite de Sitter radius,  but according to assumption  the correlation between $\CT$ and $\Pi_-$ does not go to zero;
 $\langle C  \  \Pi_-   \ra$ tends to a finite limit (see the introduction section \ref{intro}). Therefore cluster decomposition is violated in the bulk theory, signaling a spontaneous breaking of $\CT.$ 
 
 Simply stated, the existence of a tiny time-oriented feature, arbitrarily far from the laboratory  near the horizon will set off an instability that leads to a non-zero expectation value for $C,$ consistent with the non-zero value in the semiclassical theory, and cluster decomposition is restored\footnote{While this eliminates the paradox it still remains to calculate the values of the correlations in a specific  holographic theory such as \dk \ and show that they agree with the semiclassical theory.}.

 One feature of SSB is that the instability leading to a properly clustering state can be kicked off by a very weak perturbation. In fact almost any perturbation will work and lead to the same global behavior. This meshes well with the literature in which the observer  can be almost anything from a galaxy to a few low energy photons.

\subsection{The Puzzle}
Consider a bulk operator $A(t)$ which for definiteness is located in the laboratory:  the region near, but not too near the horizon. $A$ is Hermitian and time-reversal symmetric,
\bea 
A \eq A^{\dag} \cr \cr
\CT A(t) \CT &=& A(-t).
\label{Aherm}
\eea 
The operator
\be 
C = [A(-s), A(s)]
\label{C(t)}
\ee
is $\CT$ odd for any value of $s.$

The expectation value of $C$ vanishes in the maximally mixed state,
\be 
\Tr C = \Tr [A(-s), A(s)] =0.
\label{C=0}
\ee
The reason is just that the trace of a commutator vanishes.

Equation \eqref{C=0} has presented a puzzle for some time \cite{Susskind:2023rxm}, the reason being that it is not zero in the semiclassical theory of
 \dss.  The quantity $\langle C \ra $ is equal to the imaginary part of the
 correlation function $\langle A(-t) A(t)\ra.$  When calculated in QFT in a de
   Sitter background the imaginary part of two-point functions is not zero\footnote{At this point the reader may want to review sections  \ref{system} and \ref{clock}.}

This is one of those puzzles whose resolution requires the existence of a  time-oriented clock---let's say a FGC.
While \eqref{C=0} is correct the presence of a FGC breaks $\CT$ and allows a non-vanishing value of $\langle C \ra.$ To see this,   observe that both  $C$ and $\Pi_-$ are odd under time reversal so that $C\Pi_-$ is even. This allows 
\be 
\langle C  \Pi_-   \ra  = \Tr C  \Pi_-  \neq 0.
\label{neq0}
\ee

 These  observations:
  the need to measure time by a physical FGC,
but  that is not enough. In order that the clock can be arbitrarily far from the location of  $C$ and \eqref{neq0}  remain finite there must be a spontaneous breaking of 
 $\CT$ in the maximally mixed state. 
 
  Physical clocks and spontaneous breaking of $\CT$   can potentially     resolve the paradox of \cite{Susskind:2023rxm}:
the existence of a tiny time-oriented feature arbitrarily far from the laboratory  (near the horizon) will set off an instability that leads to a non-zero expectation value for $C,$ consistent with  the semiclassical theory.

\section{Gauging $\CT$ and Dressing C}

In section \ref{gauge} I discussed gauging $\CZ_2$ as a way to restore cluster decomposition. In \dss \ this may be much more natural than adding a small symmetry breaking to the holographic Hamiltonian. Harlow and Numasawa
\cite{Harlow:2023hjb} have made arguments that $\CT$ is a gauge symmetry
which I find compelling. In the $\CZ_2$ case we constructed gauge invariant operators by dressing to the ``observer,"
$$\bar{Z}_i = Z_i Z_p,$$
which rendered them gauge invariant and properly clustering. 
While $Z_i$  is odd under $K$ and is therefore not gauge invariant, $\bar{Z}_i$
is gauge invariant.

The analog of $Z_i$ is the order parameter $C.$ Like $Z_i$ it is odd (under
$\CT$)  but it can be dressed to the clock:
\be 
\bar{C}= C \Pi_-
\ee
$\bar{C}$ is the dressed version of $C;$ it is $\CT$-invariant.
Unlike correlations of $C,$  correlations of  $\bar{C}$ do cluster properly.

\bn

If, following \cite{Harlow:2023hjb} we take $\CT$ to be a gauge symmetry then $C$ is not gauge invariant,  but the dressed operator $\bar{C}$ is.

\section{Another Viewpoint} \label{vpoint}

It is interesting to consider the flat-space limit in which the de Sitter radius  $\ell \to \infty,$ and the pode moves infinitely far from the horizon. One can take the flat-space limit in two ways. One way is to focus attention on the near horizon region which includes
everything whose proper distance from the horizon remains finite in the limit $\ell \to \infty.$ This is the limit in which the static patch tends to Rindler space and the pode disappears to infinity.

The other way to take the limit is  by focusing attention on the pode while the horizon moves off to infinity. The pode just becomes the origin of Minkowski space.

So far I have thought of the observer  as living at the pode and the laboratory including $C$ as being near (but not too near) the horizon. This
seems strange because one would expect that when the observer moves off to infinity its presence should become irrelevant to the physics in the laboratory. The resolution of this puzzle---SSB---is what this paper is all about.
But it is also  interesting to center the the laboratory at the pode and
let the observer follow the horizon (or stretched horizon).

In this latter viewpoint the clock is a tiny out-of-equilibrium feature near the horizon  as long as it breaks time-reversal symmetry. (Breaking $\CT$ is very generic; not breaking it  requires fine-tuning.)

Consistency requires  the following conjecture to be true: 

\it
\bn
Conditioning on such a small horizon asymmetry 
 correlation functions near the pode, calculated in the holographic theory, agree  with their semi-classical counterparts.

\rm

\bn

I emphasize that this is a conjecture; I don't see any obstacle to it being true but it needs confirmation.

\section{Conclusion}

Depending on one's definition of observer, the claim that quantum de Sitter space is consistent only if there is an observer is either wrong, or the definition of observer is so broad that it means almost anything. What is true is that the    $\CT$ symmetric vacuum is unstable and only the feeblest perturbation will suffice to kick it into a stable spontaneously broken state. That perturbation could be the existence of a particle detector, a gravitationally bound galaxy,
a tiny black hole or a couple of photons of energy $T_{gh}$ as long as it is not exactly $\CT$-symmetric. For all but the last case the probability for the existence of such an object is exponentially
 small,
 $$\sim  \exp{\lf -\frac{M}{T_{gh}}  \rg} =
        \exp{\lf  -2\pi \ell M  \rg}$$
 but for the last case where the energy of the feature is $\sim T_{gh}$ the probability is order unity. One can summarize  this with the phrase:
Features are common; Observers are rare.
 
 In fact it is probably true that the theory without a feature is inconsistent; a state with a feeble feature is not orthogonal to the maximally mixed state. Projecting onto a state in which every conceivable feature is absent probably leaves nothing. But states without observers are consistent. If for example an observer means anything with mass greater than the Planck scale
 the probability to not have an observer is close to one,
  $$\text{Prob for no observer} =1- \left( e^{- \frac{ M_{planck} }{T_{gh}}}\right)$$
By contrast, the probability to not have an feature of any kind may be zero.

 Finally we can ask about stability of the vacuum: does it require observers or to put it another way is the Hartle-Hawking vacuum unstable. Its holographic dual---the maximally mixed state is unstable. The answer, based on Euclidean gravitational path integration  is yes, the HH state is unstable. The argument was originally due to Polchinski  \cite{Polchinski:1988ua}. Recently it was revisited in  \cite{Ivo:2025yek} and \cite{Chen:2025jqm}. The conclusion of these later papers is that the presence of an observer in the path integral is sufficient to stabilize the vacuum. It is likely that a feeble feature will also suffice.
 
 Whether the Polchinski instability is related in any way to the SSB instability discussed in this paper is an open question. Both instabilities are eliminated by the introduction of an observer but the Polchinski instability does not obviously have anything to do with time-reversal. 

I want to emphasize a point made in \cite{Chandrasekaran:2022cip}.
It is not enough to project onto states with an clock---FGC or BGC---to reproduce semiclassical correlations. It is also necessary to replace the ideal time $t$ by the clock time $\tau$ as the argument of the correlation functions.

\section*{Acknowledgement}

I have benefited from many discussions with Ying Zhao, Daniel Harlow and Edward Witten.

	\end{document}